\documentclass[manuscript, screen, nonacm]{acmart}

\usepackage{comment}

\AtBeginDocument{%
  \providecommand\BibTeX{{%
    \normalfont B\kern-0.5em{\scshape i\kern-0.25em b}\kern-0.8em\TeX}}}

\setcopyright{acmlicensed}
\copyrightyear{2020}
\acmYear{2020}
\acmDOI{10.1145/1122445.1122456}

\acmConference[NordiCHI '20]{NordiCHI '20}{October 25--29, 2020}{Tallinn, Estonia}
\acmBooktitle{NordiCHI '20,
October 25--29, 2020, Tallinn, Estonia}
\acmPrice{15.00}
\acmISBN{978-1-4503-XXXX-X/18/06}

\usepackage{framed}
\usepackage{pifont}

\usepackage{subcaption}
\usepackage{wrapfig}
\usepackage{xcolor}

\begin{document}

\title{Visualising COVID-19 Research}

  \maketitle

\vspace{-1cm}

\textbf{Pierre Le Bras, Azimeh Gharavi, David A. Robb, Ana F. Vidal, Stefano Padilla, and Mike J. Chantler}
\newline
Strategic Futures Laboratory, School of Mathematical and Computer Sciences, Heriot-Watt University, Edinburgh, UK. 
\newline
{\color{ACMDarkBlue}\{ P.Le\_Bras, AG72, D.A.Robb, AF69, S.Padilla, M.J.Chantler \}@hw.ac.uk}

\renewcommand{\shortauthors}{Strategic Futures Lab, Heriot-Watt University, Edinburgh}


\begin{center}12 May 2020 \end{center}


\newcommand{\abs}{
    The world has seen in 2020 an unprecedented global outbreak of SARS-CoV-2, a new strain of coronavirus, causing the COVID-19 pandemic, and radically changing our lives and work conditions. Many scientists are working tirelessly to find a treatment and a possible vaccine. Furthermore, governments, scientific institutions and companies are acting quickly to make resources available, including funds and the opening of large-volume data repositories, to accelerate innovation and discovery aimed at solving this pandemic. In this paper, we develop a novel automated theme-based visualisation method, combining advanced data modelling of large corpora, information mapping and trend analysis, to provide a top-down and bottom-up browsing and search interface for quick discovery of topics and research resources. We apply this method on two recently released publications datasets (Dimensions' COVID-19 dataset and the Allen Institute for AI's CORD-19). The results reveal intriguing information including increased efforts in topics such as social distancing; cross-domain initiatives (e.g. mental health and education); evolving research in medical topics; and the unfolding trajectory of the virus in different territories through publications. The results also demonstrate the need to quickly and automatically enable search and browsing of large corpora. We believe our methodology will improve future large volume visualisation and discovery systems but also hope our visualisation interfaces will currently aid scientists, researchers, and the general public to tackle the numerous issues in the fight against the COVID-19 pandemic.
}

\vspace{-.4cm}
\section*{Abstract}
\abs

\vspace{0.1cm}

\noindent \textbf{Keywords}: COVID-19; Coronavirus; SARS-COV-2; Information Visualisation; Research Corpora; Topic Modelling.

\vfill 

\begin{figure}[h]
    \centering
    \includegraphics[width=1.0\columnwidth]{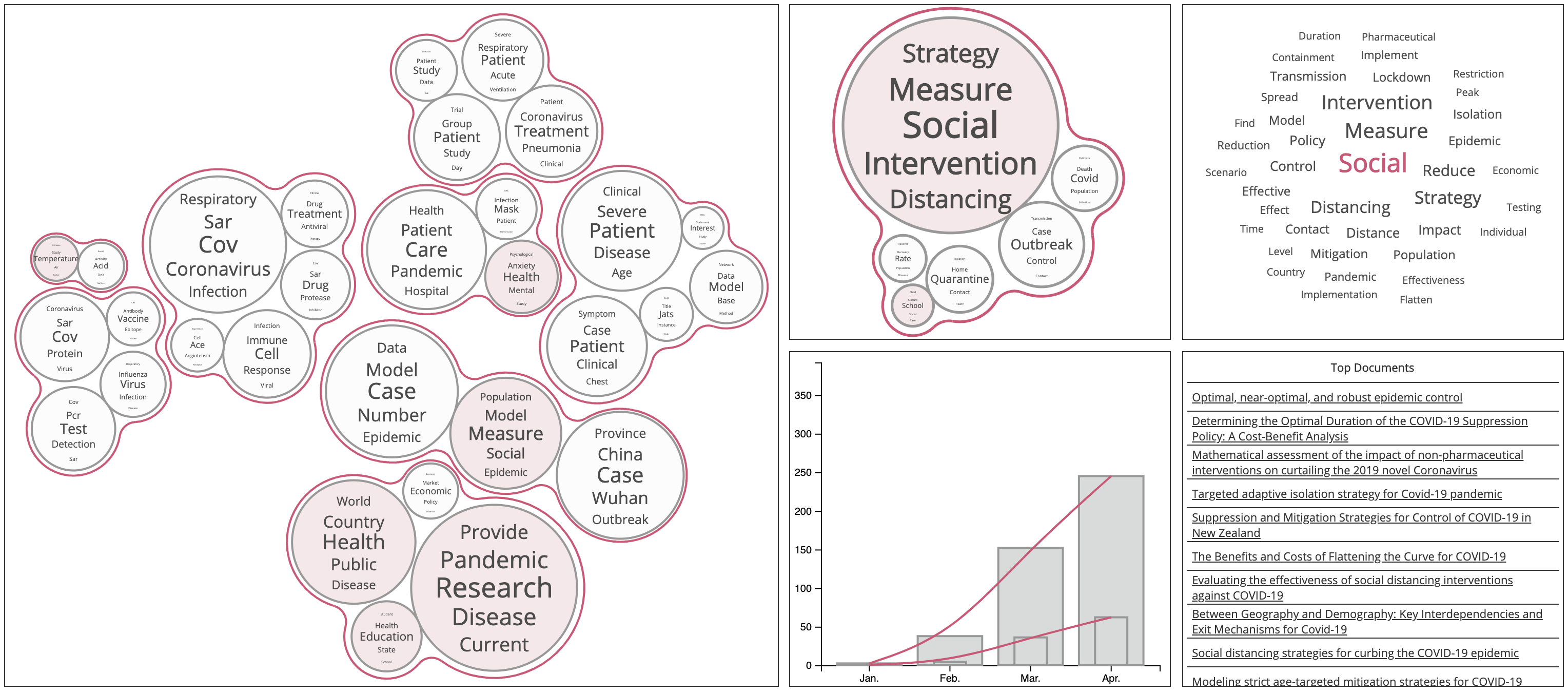}
    \vspace{-0.6cm}
    \caption{Our online system demonstrates the hierarchical topic visualisation of the Dimensions COVID-19 dataset. The principal panel on the left shows the at-a-glance overview of all the resources in the dataset. The panels on the right visualise the sub-topics when selecting a main topic, the topic descriptions, the trend analysis, and the drill-down to available resources. Moreover, the system includes a search capability for the bottom-up discovery of known areas and support for advanced users. The interactive visualisation is available on our research lab pages: \href{http://strategicfutures.org/TopicMaps/COVID-19/dimensions.html}{http://strategicfutures.org/TopicMaps/COVID-19/dimensions.html}.}~\label{fig:app}
\end{figure}

\section{Introduction}

As the scientific community grapples with the search for solutions to the COVID-19 pandemic, it also needs to accelerate the pace of innovation \cite{ahamed2020information}. Globally, research funding organisations are seeking to further drive an acceleration in research with additional short duration calls, schemes to re-purpose existing funding and opening research resources, e.g. \cite{mrc-rolling-call, UKRI-opencall}. With this need for faster-paced research comes a need for timely processing, assimilation and appreciation of the continually growing and evolving body of research literature that is originating rapidly from the global research effort. \newline 

Recent studies have examined the growing literature on COVID-19 and employed various methods and techniques to analyse it. Haghani et al. \cite{Haghani20} use bibliometric analysis with heat maps, pie charts, and bar graphs to describe COVID-19 research areas and their relative importance. Others combine bibliometrics with text mining. Hossain \cite{Hossain20} does this, using bibliometrics combined with text mining for word co-occurrence and factorial analysis of the top keywords visualised in network diagrams and dendrograms. Similarly (to Hossain) Aguado-Cort\'es \& Casta\~no \cite{Aguado-Cortes20} use bibliometrics and concurrence of keywords with network diagrams to visualise the analysis. Fister et al. \cite{Fister20} use association rule text mining \cite{Agrawal93}, then analyse word relationships and employ bar chart and word cloud visualisations. Wang et al. \cite{Wang20} create an application which uses distantly supervised named entity recognition \cite{Wang19} and facilitates text queries. It visualises query results with a doughnut chart. One study by Domingo-Fernandez et al. \cite{Domingo-Fernandez20} took a subset of the literature focusing on drug targets, generated a network graph by manually annotating evidence text from the corpus with Biological Expression Language (BEL) and explored the network graph with web applications built for the task. 
Lastly, Ahamed \& Samad \cite{ahamed2020information} use a method of topic analysis where topics are identified using betweenness centrality measurement \cite{Freeman77}. Subgraphs are then generated for the influential topics, and they examine several topics in detail through these subgraphs.\newline

In this paper, we introduce a COVID-19 research knowledge visualisation designed to help researchers and research strategists to get an at-a-glance overview of the research landscape (Figure~\ref{fig:app}). This overview is laid out by topics, and it is valuable to (a) identify connections between research areas, (b) recognise trends over time by visualising research volume in specific topics, and (c) find available resources from these large volume datasets of research. We believe our approach to visualise the relationships between the concepts in the COVID-19 research offers a more suitable pipeline for a frequently updateable visualisation.\newline

We use Latent Dirichlet Allocation (LDA) \cite{LDA} to build a topic model from the research corpus titles and abstracts in a similar manner to Padilla et al. \cite{padilla14}. LDA allows control of the number of topics and thus the ability to choose a suitable level of abstraction for both overview and detailed sub-topics. The topics can be interrogated intuitively to reveal more detail, and further visualised with a word cloud and a trend chart showing how the volume of research in that topic has changed over time. Our method which we describe in Section 2, uses a pipeline perfected over recent years allowing a  rapid and efficient generation of a newly updated visualisation as the corpus grows and develops, thus allowing efficient and timely updating of the visualisation.\newline

We first describe, in this paper, the methodology we use to create our visualisation for analysing the COVID-19 research landscape and present the open visualisation itself, providing the web address where it can be accessed. Then with the aid of screenshots from our interactive visualisations of two research corpora, we describe four research trends which illustrate the benefits that our techniques bring to the understanding of COVID-19 and Coronavirus research. Finally, we conclude by summarising the benefits of our techniques and our new COVID-19 research landscape analysis and visualisation tool.

~\\
In summary, the contributions of this paper are: 
\begin{enumerate}
\item We explore a novel automated theme-based visualisation methodology combined with data modelling of large corpora of COVID-19 resources.
\item We develop COVID-19 research information mapping and trend analysis to provide a top-down and bottom-up browsing and search interface for quick discovery of topics and resources.
\item We reveal interesting information from open COVID-19 research datasets using our visualisation and method. 
\item We provide an open interface for discovering COVID-19 research with the aim to aid in solving various issues of the current pandemic.
\end{enumerate}

\section{Methodology}

\begin{figure}[t]
  \centering
  \includegraphics[width=1\columnwidth]{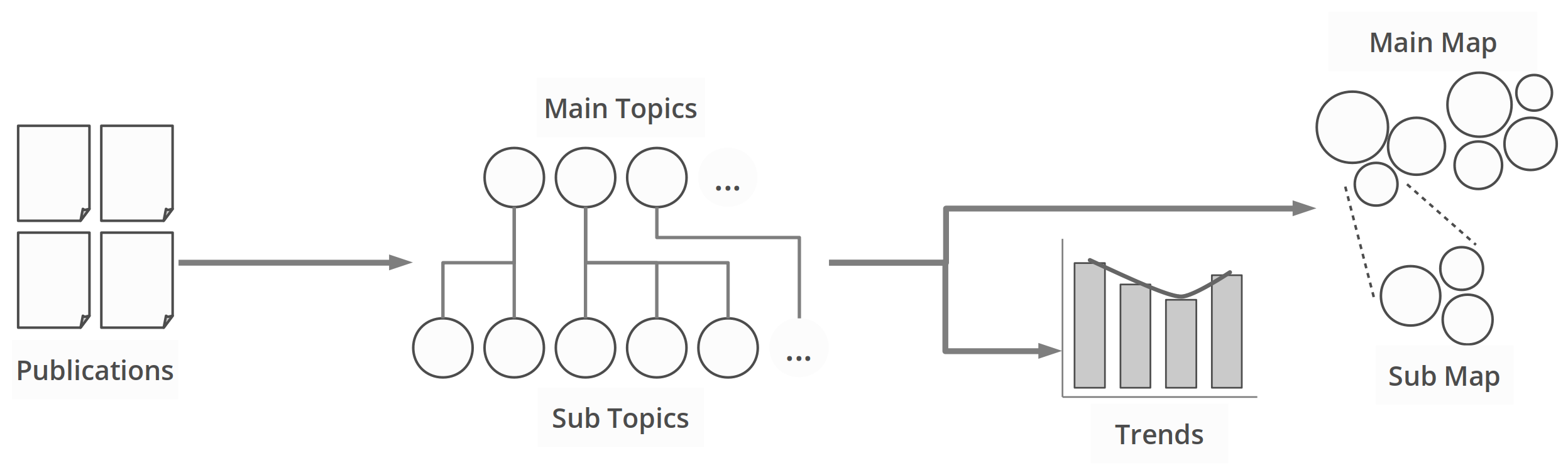}
  \vspace{-0.7cm}
  \caption{Our approach to visualising research publications. Starting from the publication titles and abstracts, we generate a hierarchical topic model: a main topic model with tens of topics, and a sub-topic model with hundreds of topics. Each sub-topic is assigned to one main topic based on their document vector similarity. Topic trends are then extracted, and topics are mapped into bubble maps: the top-level map containing the main topics, and tens of sub-maps, one for each main topic, containing sub-topics.}~\label{fig:methodology}
  \vspace{-0.5cm}
\end{figure}

In this section, we present our methodology for visualising research information of a large volume. In particular, we use topic modelling to abstract thousands of research documents into a smaller hierarchical set of themes. We then estimate the trends of these topics and group these into simplified bubble treemaps to create semantic overviews. Figure~\ref{fig:methodology} illustrates this automated process.

\subsection{Datasets}

Our methods and visualisations are automatically generated from research corpora. This automated approach is particularly suited for rapidly evolving knowledge datasets as it requires little oversight, manipulation, and can avoid common delays frequently encountered in manual classifications and processes.
\newline

\begin{figure}
\begin{subfigure}[t]{.4\textwidth}
  \centering
  \includegraphics[height=5cm]{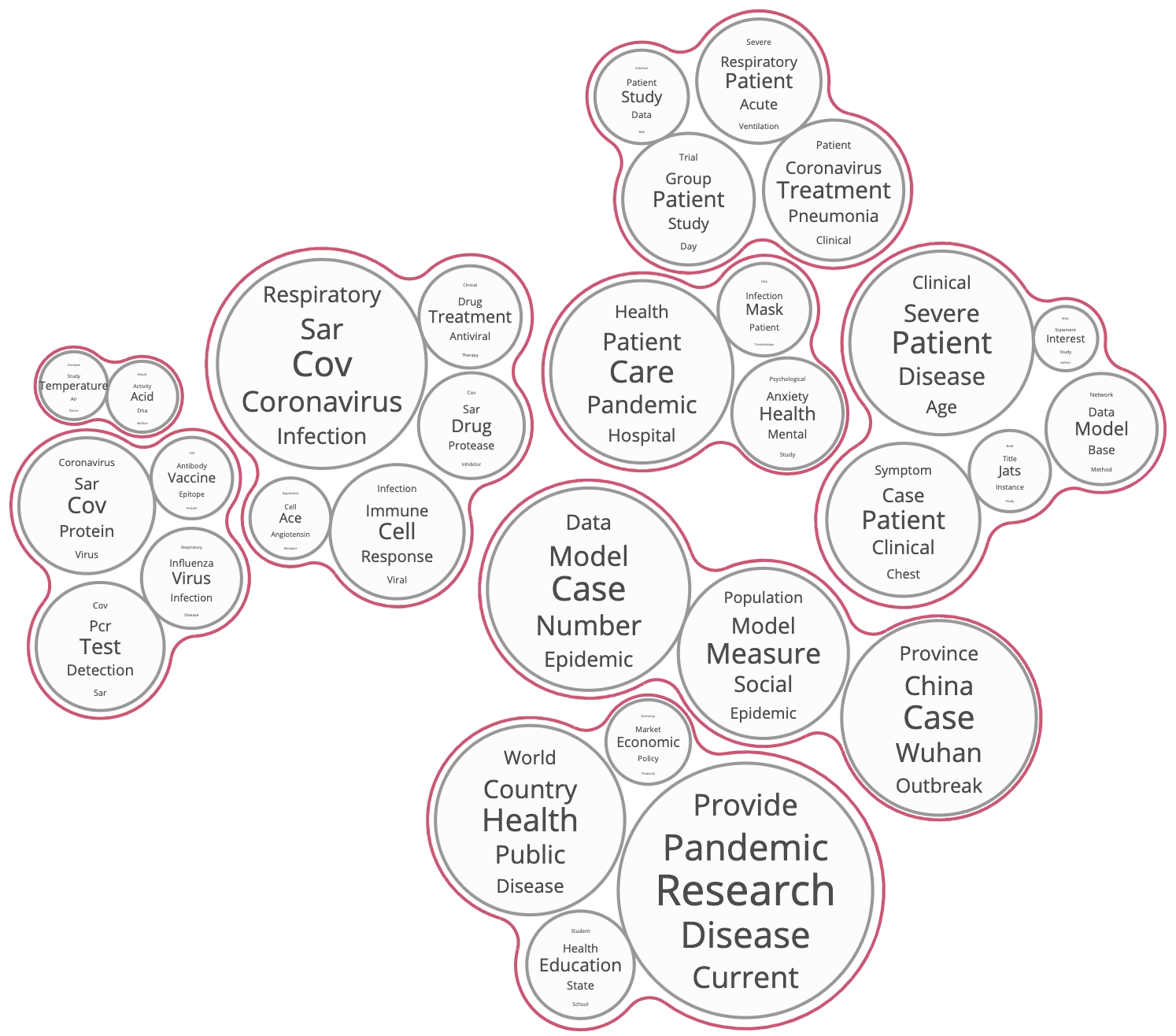}
\end{subfigure}
\begin{subfigure}[t]{.5\textwidth}
  \centering
  \includegraphics[height=5cm]{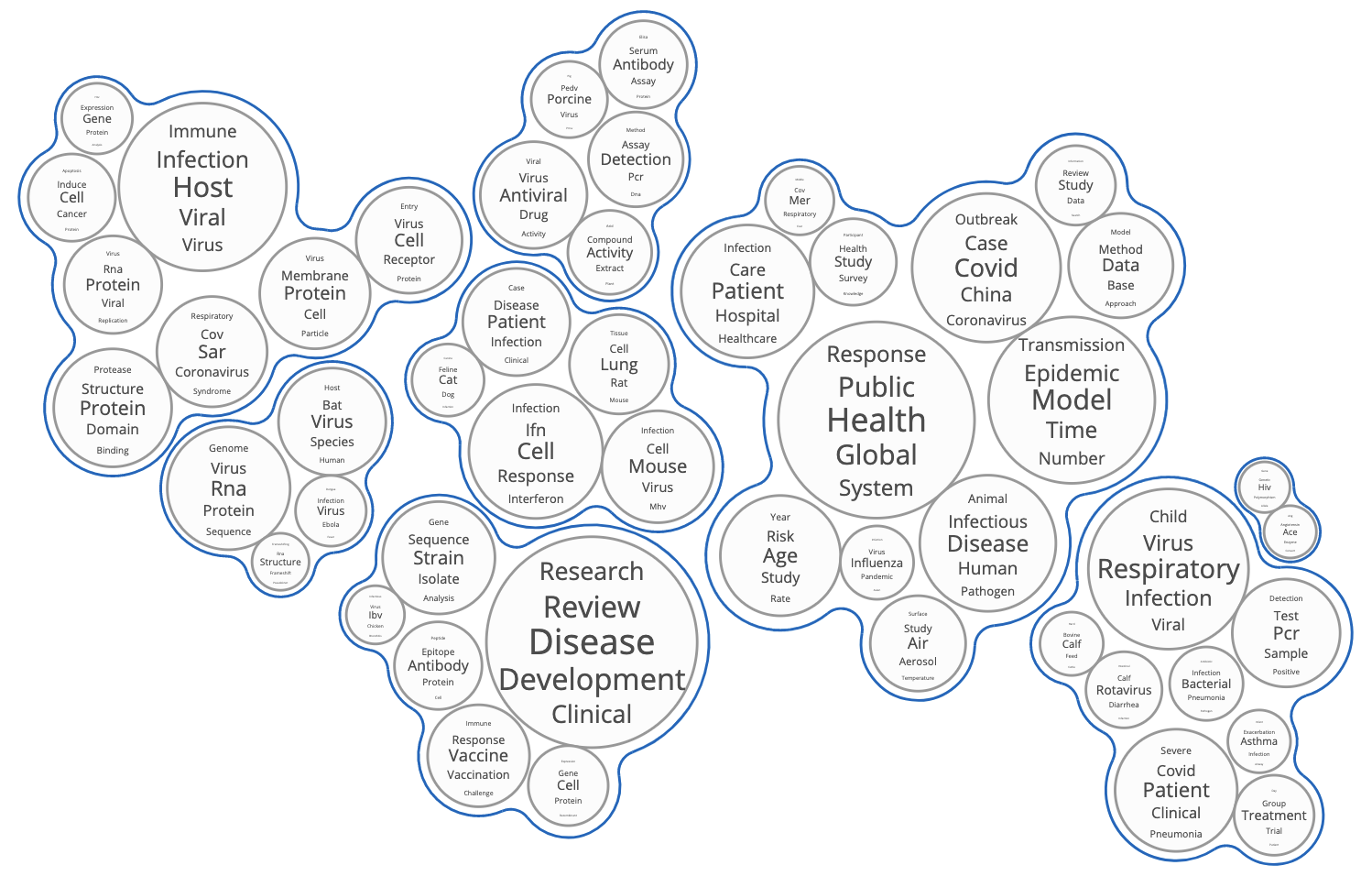}
\end{subfigure}
\vspace{-0.3cm}
\caption{The two datasets chosen for the visualisation system. On the left and in red is the representation of the Dimensions COVID-19 research dataset, and on the right in blue is the representation of the COVID-19 Open Research Dataset (CORD-19).}~\label{fig:dataset}
\vspace{-0.5cm}
\end{figure}

We have chosen to first work on the Dimensions COVID-19 research dataset \cite{dimensionsDataset} as it curates, from multiple sources, the details of publications submitted in the last four months investigating the COVID-19 outbreak. We extracted the publications' title and abstract; from our experience, we have found this information is sufficient for generating meaningful representative topics (Figure~\ref{fig:dataset} in red). In total, $17,015$ publications were retrieved from the 16\textsuperscript{th} version of this dataset, dated from the 23\textsuperscript{rd} of April  2020. Moreover, our process can scale to any corpus input, and we have also chosen to separately visualise the COVID-19 Open Research Dataset (CORD-19) \cite{cord19}, as it adds a historical perspective on coronaviruses research (Figure~\ref{fig:dataset} in blue). In total, $59,875$ publications' titles and abstracts were retrieved from the 12\textsuperscript{th} version of this dataset, dated from the 4\textsuperscript{th} of May 2020. The rest of the process is entirely automated using an algorithmic pipeline which consists of three main phases: \textit{modelling topics}, \textit{estimating trends}, and \textit{mapping topics}. These stages are explained in detail in the next subsections.

\subsection{Modelling Topics}

We produced a two-level hierarchical topic model to allow for fast visual discovery, cognitive processing, and cognitive retention of the topics in users. We did so by separately modelling 30 topics and 200 topics from the same publication data (Dimensions), using MALLET's implementation of Gibbs Sampling for Latent Dirichlet Allocation (LDA) \cite{LDA,GibbsSampling,MALLET}. Care was taken to process the datasets appropriately; this includes removing general stop words, lemmatising terms and choosing the most beneficial parameters for the implementation. To do so, we used knowledge from previous work \cite{LeBras2018} and followed recommendations from Boyd-Graber et al. \cite{Boyd2014Care}.  Each sub-topic (200) was then assigned to the one main topic (30) with the least cosine distance between their document vectors. Because the CORD-19 dataset contained more historical publications (e.g. SARS and MERS), we used larger numbers of topics to model: 50 main topics and 400 sub-topics.

\subsection{Estimating Trends}

We believe it is valuable to understand the evolution of the themes. While topic modelling offers the ability to abstract thousands of documents into a more digestible set of themes, it is also essential to represent the time trajectory of the resources to enable researchers to compare current trends of research quickly. Using the publication date information, and the topic weights in each publication, we were able to construct the distribution of each topic over time. This distribution details, per topic and across dates, the sum of this topic's weights for publications of that date\footnote{It was noted that, in the Dimensions dataset, some publications where scrapped with only the information ``2020''. As a result, they were defaulted to the 1\textsuperscript{st} of January, making this date showing an unusual volume of publications. To simplify the reading of trend charts, we have excluded this date.}. These trend charts are displayed when a user interacts with a topic as shown in Figure~\ref{fig:app} for the \textit{Social-Measure-Intervention} sub-topic.

\subsection{Mapping Topics}

We use a novel implementation of bubble treemaps to visualise our sets of topics \cite{BubbleTreemap}. We choose this representation as it allows us to show (a) the topics relative similarities with the placement of bubbles, (b) the importance of the topic with the size of bubbles, and (c) clusters of topics to facilitate cognitive recognition and memory. We first computed the hierarchical topic similarities using the cosine distance between the topics' document vectors, and agglomerative clustering with a complete linkage criterion \cite{AgglClustering}. Then, the bubble treemap placement algorithm on this hierarchy allowed us to position similar topics next to one another. We used the sum of each topic weights across publications to set the bubble sizes, showing the topics relative importance in the dataset. To represent our two-level topic hierarchy, we first mapped the main topic model, the resulting visualisations for both datasets are shown in Figure~\ref{fig:dataset}. Then, for each main topic, we grouped the associated sub-topics and mapped these groups independently from one another. It resulted as a set of sub-maps, each accessible from their main topic. Finally, this automated methodology allowed us to quickly abstract and visualise the content of thousands of research publications on COVID-19 and coronaviruses. In the next sections, we present these visualisations and highlight compelling findings within them.  

\section{Overview of COVID-19 Research}

 \begin{figure}[t]
  \centering
  \includegraphics[height=7cm]{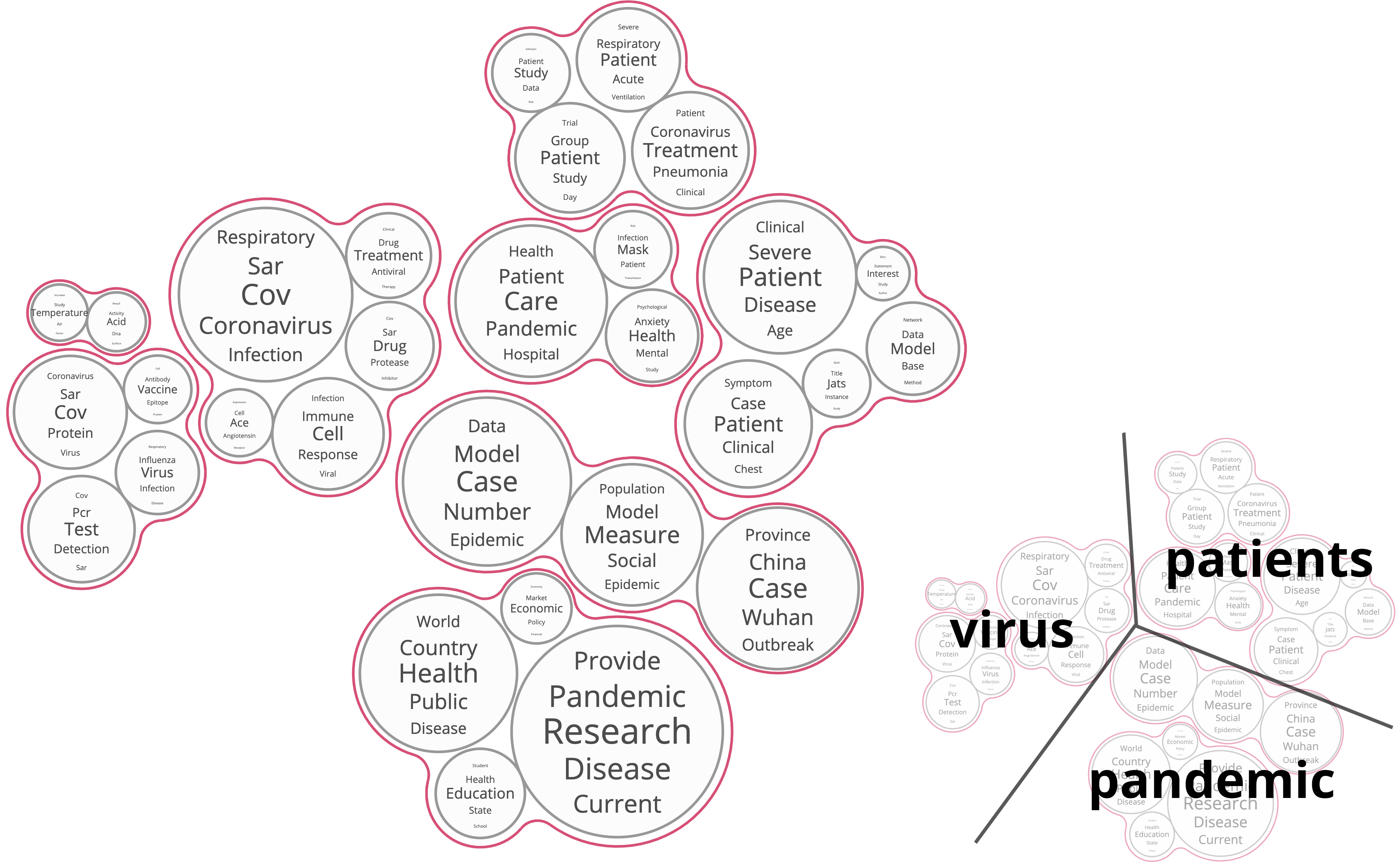}
  \vspace{-0.3cm}
  \caption{Top-level overview of the COVID-19 research publications in the Dimensions dataset. Each bubble contains the top labels of the extracted topics, and the bubble size show the proportion of the topic. The placement and clustering of bubbles depict the relative similarities between topics. We have found that three significant regions emerged: \textit{Patient Care} at the top right-hand side; \textit{Virus and Micro-Biology} at the left-hand side; and \textit{Pandemic} at the bottom.}~\label{fig:dimensions_map}
  \vspace{-0.7cm}
\end{figure}

We present the bubble map overview of COVID-19 Research in Figure~\ref{fig:dimensions_map}. Because it focuses specifically on the latest publications regarding COVID-19, this map uses the Dimensions datasets described in the previous section. While this visualisation depicts eight clusters\footnote{We have selected this number as it compromised on quantity and quality of clusters.}, we can identify three major regions in this map

\begin{enumerate}
    \item On the top right-hand side corner, we see topics dealing with \textit{patients}, \textit{treatment}, and \textit{care}. These cover domains such as common symptoms and clinical trials of possible treatments for COVID-19, modelling of the disease, hospital management, transmission research, and mental health concerns.
    \item On the left-hand side of the map, the topics seem to represent domains like \textit{virology} and \textit{micro-biology}. It is in this region that we find research investigating the actual virus SARS-COV-2, its genome, testing, drugs effects, and possible vaccines.
    \item At the bottom of the map, we find bigger topics, all with regards to the \textit{pandemic} and its consequences. This includes modelling the pandemic, and discussion about its economic political, and societal impacts.
\end{enumerate}

In the next section, we present four detailed analyses indicative of the emerging and rapidly changing focuses this pandemic has unfold.

\section{Discussion}

Virology, vaccine, antiviral, health, and treatment research gratefully constitute the core of the scientific response to COVID-19. This unprecedented situation has, however, also seen the emergence of research in many other fields. By modelling these topics, and mapping them in semantic layouts, we were able to discover exciting themes, navigate this research and highlight interesting trends. 

\subsection{Unprecedented Social Distancing}

\begin{figure}
\hfill
\begin{subfigure}[t]{.28\textwidth}
    \centering
    \includegraphics[width=\columnwidth]{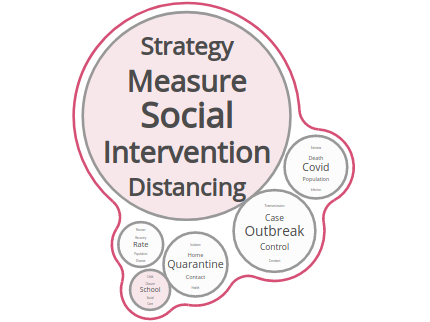}
    \caption{In the main topic \textit{Measure-Model-Social}, the sub-topic \textit{Social-Measure-Intervention} predominates.}~\label{fig:social_d_submap}
\end{subfigure}
\hfill
\begin{subfigure}[t]{.28\textwidth}
    \centering
    \includegraphics[width=\columnwidth]{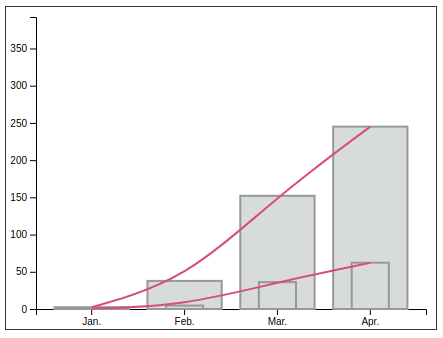}
    \caption{Both main and sub-topics about social distancing show a steep increase in terms of volume of research, starting from February 2020.}~\label{fig:social_d_trend}
\end{subfigure}
\hfill
\begin{subfigure}[t]{.28\textwidth}
    \centering
    \includegraphics[width=\columnwidth]{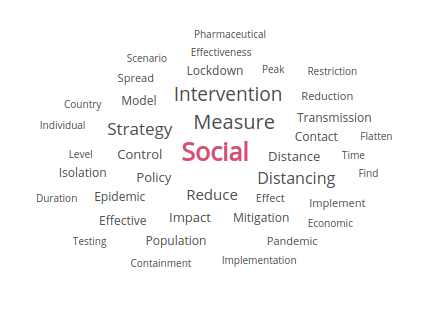}
    \caption{The word cloud for the sub-topic \textit{Social-Measure-Intervention} informs us on the vocabulary used to describe this issue.}~\label{fig:social_d_wordcloud}
\end{subfigure}
\hfill
\vspace{-.5cm}
\caption{In the Dimensions dataset, social distancing is shown as a trending research topic.}~\label{fig:social}
\vspace{-.7cm}
\end{figure}

Three months after first being reported, by March 2020, this pandemic has launched an unprecedented global response to slow down its spread. For many countries, this response meant limiting human contacts (and possible transmission) to the maximum, hence introducing \textit{social distancing}. We examine the importance of this issue in research, as illustrated in Figure~\ref{fig:social_d_submap}, which shows the sub-topic \textit{Social-Measure-Intervention} predominating other sub-topics in the main topic \textit{Measure-Model-Social}. We can also measure the rapid ascent of this theme from February, to March and April 2020 in Figure~\ref{fig:social_d_trend}, corresponding to the point when many countries followed China's suite and put lockdown measures in action.
\newline

\begin{figure}
\hfill
\begin{subfigure}[t]{.28\textwidth}
    \centering
    \includegraphics[width=\columnwidth]{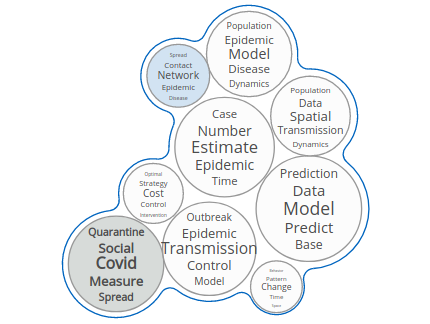}
    \caption{We only find mention of social distancing in the topic \textit{Model-Epidemic-Time}, and it is only next to \textit{COVID}.}~\label{fig:social_k_submap}
\end{subfigure}
\hfill
\begin{subfigure}[t]{.28\textwidth}
    \centering
    \includegraphics[width=\columnwidth]{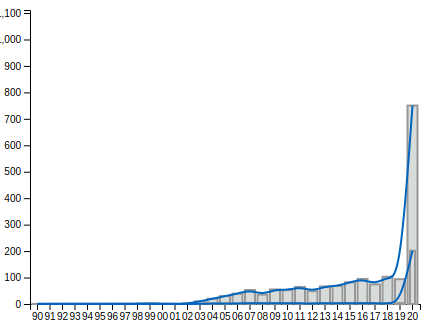}
    \caption{The historical trend of social distancing (lower line) only seems to show an increase in 2020.}~\label{fig:social_k_trend}
\end{subfigure}
\hfill
\begin{subfigure}[t]{.28\textwidth}
    \centering
    \includegraphics[width=\columnwidth]{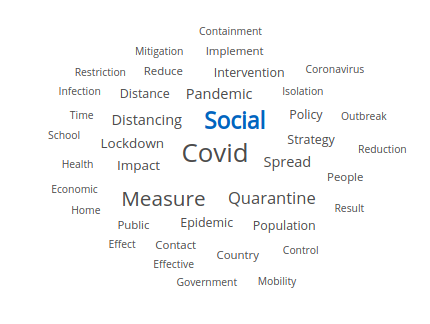}
    \caption{Although we could expect the topic trend to be caused by the term \textit{COVID}, we see that most labels of that topic focus on distancing.}~\label{fig:social_k_wordcloud}
\end{subfigure}
\hfill
\vspace{-.5cm}
\caption{The CORD-19 dataset interface allows us to explore a historical perspective on social distancing.}~\label{fig:social}
\vspace{-.7cm}
\end{figure}

To get a broader perspective, we have searched for similar topics in the CORD-19 dataset, which comprise older research on coronaviruses (from 1951). There, we find that social distancing is only mentioned along \textit{COVID} (Figure ~\ref{fig:social_k_submap}). Furthermore, while recent papers constitute 25\% of the whole corpus, we can see that the trend for this sub-topic only shows an increase in 2020 (Figure~\ref{fig:social_k_trend}, lower line). It could be in part due to \textit{COVID-19} being categorised in this sub-topic, however, checking the other labels in the sub-topic (Figure~\ref{fig:social_k_wordcloud}), we only find a few occurrences of COVID related terms (\textit{Spread}, \textit{Epidemic}, \textit{Pandemic}) compared to a larger number of labels describing social distancing (\textit{Measure}, \textit{Lockdown}, \textit{Quarantine}, \textit{Intervention}, \textit{Mitigation}, \textit{Mobility}). It reflects that this is the first time, in peacetime, that a lockdown of this scale has been imposed to slow the exponential growth in transmission rate.
\newline

We found that the research categorised in these topics can provide a resource to understand and optimise social distancing measures, such as (a) drawing models of the costs and benefits of these measures against their effectiveness; (b) analysing the impact of asymptomatic carriers; (c) suggesting effective exit strategies; or (d) predicting long term social and travel behaviours. Measures such as social distancing have also triggered society to adapt and find new ways of continuing its activities. We discuss these societal impacts below and how scientists address them.

\subsection{Research Responding to Society's Concerns}

\begin{figure}[t]
\hfill
\begin{subfigure}[t]{.28\textwidth}
  \centering
  \includegraphics[width=\columnwidth]{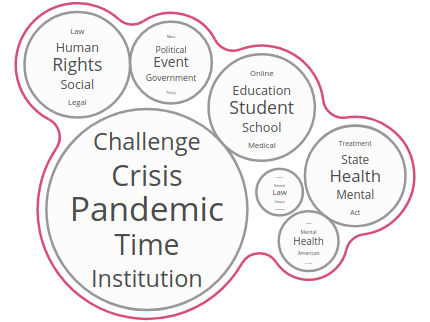}
  \caption{With crises and challenging times emerge new issues regarding rights, education, or mental health.}~\label{fig:societal_submap}
\end{subfigure}
\hfill
\begin{subfigure}[t]{.28\textwidth}
  \centering
  \includegraphics[width=\columnwidth]{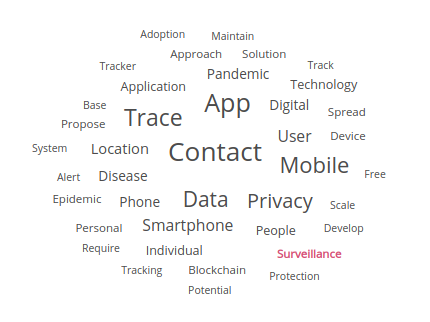}
  \caption{Contact tracing and surveillance can enable better prevention, but needs strong ethical considerations.}~\label{fig:contact_tracing}
\end{subfigure}
\hfill
\begin{subfigure}[t]{.28\textwidth}
  \centering
  \includegraphics[width=\columnwidth]{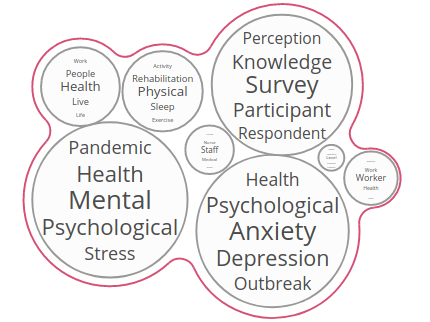}
  \caption{Mental health can be challenging in difficult times; it is therefore crucial to understand and address such issues.}~\label{fig:mental_health}
\end{subfigure}
\hfill
\vspace{-.5cm}
\caption{With Topic Modelling we can explore the emergence of new concerns and research with regards to the impacts of this pandemic on society.}~\label{fig:social_panels}
\vspace{-.6cm}
\end{figure}

Topic modelling is a technique that allows us to analyse large amounts of information to get an overview of all emerging issues in text corpora. As such, we can gain access to often-overlooked topics or themes. In the Dimensions dataset, for example, we can find a group of sub-topics, under \textit{Education-Health-State}, all with regards to the societal impacts of this pandemic. We show these topics in Figure~\ref{fig:societal_submap}. Although a large portion of publications describes a \textit{Pandemic}, \textit{Crisis}, or \textit{Challenge}, we find that they also address issues such as public health and individual rights, mental health, education, and economy. 
\newline

Responding to any outbreak requires policymakers to modify or enact new public health laws. The COVID-19 pandemic, however, has seen an even greater and more urgent public health response due to the high transmission rate of the virus. Research has, therefore, emerged to investigate the impact of these measures on individual rights \cite{meier2020rights}. One common concern is individual liberty faced with forced social distancing and isolation. Another is medical privacy versus the prevention of virus spread, i.e. by reporting and disclosing patients' name to their employers, or large-scale surveillance and contact tracing. This latter example is even large enough to have its own sub-topic, where technological vocabulary meets with ethics and privacy (Figure ~\ref{fig:contact_tracing}).
\newline

As we have seen before, social distancing has been a significant consequence of this pandemic. It is therefore natural for researchers to explore the effects of such measures. We have found three major fields of research in our topic maps. The first concerns mental health, and seems to follow two paths. On the one hand, there is research on the effectiveness of new approaches for pursuing existing therapies remotely, either one-on-one or group support meetings to manage addiction. On the other hand, we recognise studies focusing on the overall population's stress, anxiety, and depression facing the pandemic, as well as concerns towards the mental strain on medical and care staff working endlessly to help patients. The importance of that field has made it one of the main topics in the Dimensions dataset, from which we show the detailed submap in Figure~\ref{fig:mental_health}.
\newline

The second addresses the new challenges for education. With schools and universities closing to protect their students and prevent the virus from spreading, teachers and administrations have to adapt to provide learning resources remotely. It is then necessary to study the effectiveness of large-scale home-schooling, as well as new online teaching methods for elementary, higher and university levels. On a more direct response to the outbreak, we have also found a shift in medical training, where advanced medical students are offered to start their internship early to help treat patient and relieve stress on the other medical staff.
\newline
 
Finally, there has also been significant concerns around the economic impacts of this pandemic, present and future. In particular, we see economists and actuarial scientists analysing its damaging effects on the world market, stock prices, countries' local economies; moreover, they are trying to provide solutions to such issues. We also found that research into modern digital currencies and banking has been carried out to understand their potential as a solution to slow and prevent the progression of the virus.

\subsection{Responding to a New Disease}

\begin{figure}
\hfill
\begin{subfigure}[t]{.28\textwidth}
  \centering
  \includegraphics[width=\columnwidth]{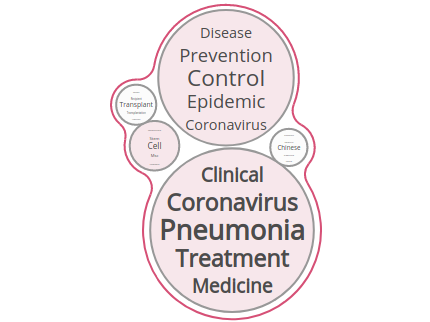}
  \caption{The sub-map of topic \textit{Treatment-Coronavirus-Pneumonia} shows how \textit{treating pneumonia} has been as critical as \textit{preventing and controlling the epidemic}.}~\label{fig:pneumonia_submap}
\end{subfigure}
\hfill
\begin{subfigure}[t]{.28\textwidth}
  \centering
  \includegraphics[width=\columnwidth]{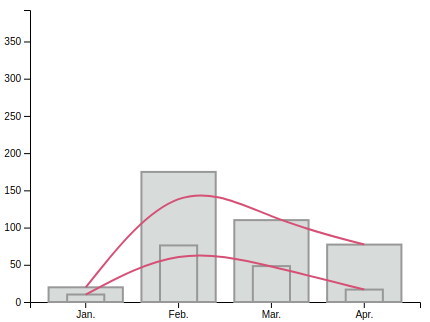}
  \caption{Looking at the trend for topics around \textit{Pneumonia}, we see it peaks in February, before slowly decreasing.}~\label{fig:pneumonia_trend}
\end{subfigure}
\hfill
\begin{subfigure}[t]{.28\textwidth}
  \centering
  \includegraphics[width=\columnwidth]{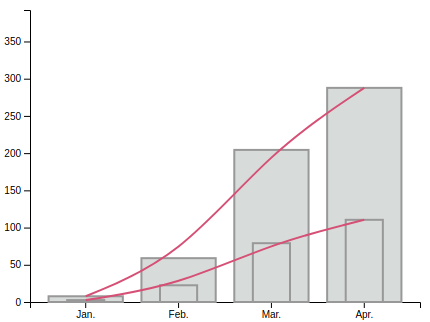}
  \caption{The trends for \textit{SARS} and \textit{COV} have only gone up, and show a steep increase after February.}~\label{fig:cov_trend}
\end{subfigure}
\hfill
\vspace{-.5cm}
\caption{Trend analysis allows us to see evolution in medical research.}~\label{fig:pneumonia_d_panels}
\vspace{-.6cm}
\end{figure}

Looking at terms such as \textit{Pneumonia} and \textit{SARS-CoV-2} probably highlights best the scientific community's response to a crisis of this sort. Looking at the topic \textit{Treatment-Coronavirus-Pneumonia} and its large sub-topic about \textit{Pneumonia} (Figure~\ref{fig:pneumonia_submap}), we find that this theme has peaked in February, only to decline in the following two months (Figure~\ref{fig:pneumonia_trend}). We suspect that it reflects the first stage of a new disease outbreak: medical staff report on the observation and study of similar symptoms rising in number of cases \cite{chahrour2020bibliometric}. Meanwhile, temporarily only known as \textit{2019-nCoV}, the virus responsible for this epidemic is sparingly used in publications. It is only by March, when it is named by the Coronaviridae Study Group (CSG) as \textit{SARS-CoV-2} \cite{sarscov2}, that we see a net increase of its mention in publication (Figure~\ref{fig:cov_trend}), taking over the initial reports on \textit{Pneumonia}. \newline

\begin{figure}[t]
\hfill
\begin{subfigure}[t]{.25\textwidth}
    \centering
    \includegraphics[width=\columnwidth]{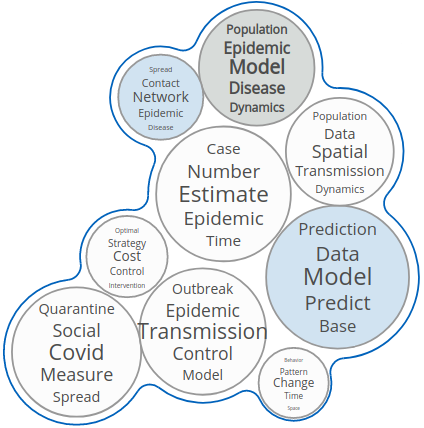}
    \caption{ The  sub-map  of  topic \textit{Epidemic-Model-Time}}~\label{fig:simulation_k_submap}
\end{subfigure}
\hfill
\begin{subfigure}[t]{.55\textwidth}
    \centering
    \includegraphics[width=\columnwidth]{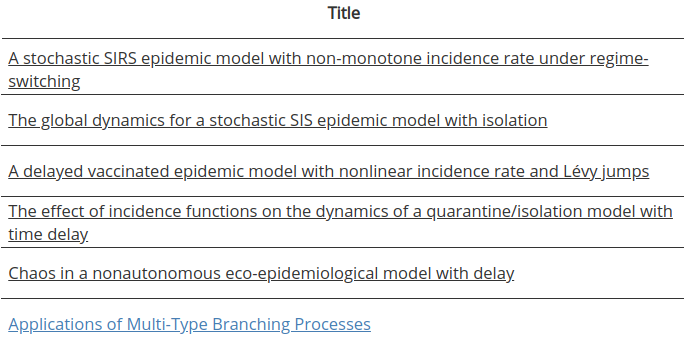}
    \caption{An excerpt from the top document list. We highlight, in blue, a relevant publication that would be difficult to discover just from its title.}~\label{fig:simulation_k_docs}
\end{subfigure}
\hfill
\vspace{-.5cm}
\caption{The CORD-19 dataset interface allows researchers to discover relevant related work that might otherwise be difficult to find from the publication titles alone. }~\label{fig:simulation}
\vspace{-.5cm}
\end{figure}

At the same time, looking at terms like \textit{Simulation} reveals the massive effort that has gone into developing better computational models to inform government policy. Many scientists have been working against the clock to use the limited available data to make predictions about the evolution of the pandemic. The interface we developed can help speed up progress by helping scientists find relevant related publications. 
In Figure~\ref{fig:simulation} we show an excerpt of the top documents for the sub-topic \textit{Data-Model-Predict} within the topic \textit{Epidemic-Model-Time} and we highlight in blue one particular publication which would be otherwise difficult to find due to the lack of epidemic-related keywords in its title.

\subsection{Tracking the International Epidemic}

\begin{figure}[t]
\hfill
\begin{subfigure}[t]{.24\textwidth}
  \centering
  \includegraphics[width=\columnwidth]{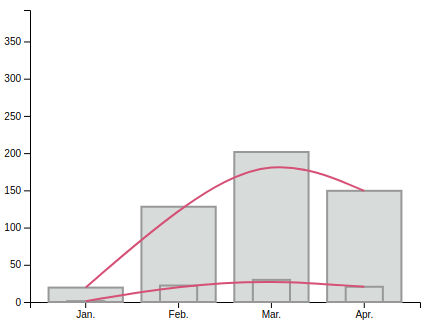}
  \caption{Publications around the outbreak in China where numerous, but only peaked in March.}~\label{fig:pandemic_china}
\end{subfigure}
\hfill
\begin{subfigure}[t]{.24\textwidth}
  \centering
  \includegraphics[width=\columnwidth]{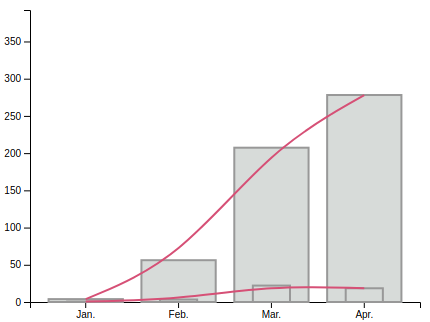}
  \caption{Publications around the outbreak in Korea, Japan, and Iran grew in March and plateaued through April (lower line).}~\label{fig:pandemic_korea}
\end{subfigure}
\hfill
\begin{subfigure}[t]{.24\textwidth}
  \centering
  \includegraphics[width=\columnwidth]{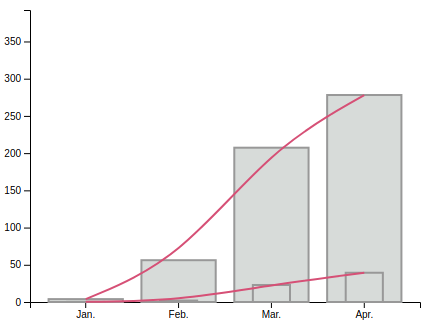}
  \caption{Publications around the outbreak in Europe (Italy, Spain, France, and Germany) have grown steadily from March into April (lower line).}~\label{fig:pandemic_europe}
\end{subfigure}
\hfill
\begin{subfigure}[t]{.24\textwidth}
  \centering
  \includegraphics[width=\columnwidth]{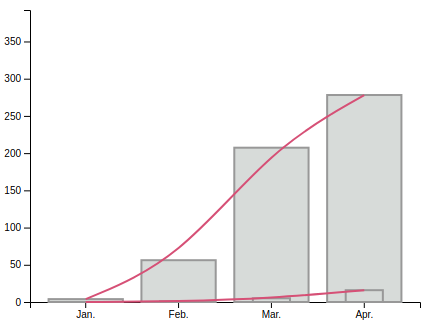}
  \caption{Publications around the outbreak in India only appeared in April (lower line).}~\label{fig:pandemic_india}
\end{subfigure}
\hfill
\vspace{-.5cm}
\caption{The trend analysis allows us to trace the virus spread through research publications (note that Fig.~\ref{fig:pandemic_china} shows both main and sub-topics about Wuhan and China; Fig.~\ref{fig:pandemic_korea}, \ref{fig:pandemic_europe}, and \ref{fig:pandemic_india} show sub-topics nested under the main topic \textit{Case-Model-Number}).}\label{fig:pandemic}
\vspace{-.3cm}
\end{figure}

A close examination of the topics in the \textit{Pandemic} region of the main topic map (see Figure~\ref{fig:dimensions_map}) allows us to visualise the trends in the volume of publication which share topics including particular countries. Upon closer inspection, we found that we can track the progress of the pandemic around the world by following the volume of publications. We show the evolution of these trends in Figure~\ref{fig:pandemic}. In Figure~\ref{fig:pandemic_china}, we see how both main and sub-topics featuring \textit{Wuhan} and \textit{China} display a trend showing publication volumes peaking in March and dropping back in April. Then, if we examine the sub-topics of the \textit{Model-Case-Number} main topic, we can see that the publication volumes trend for the sub-topic featuring \textit{Korea}, \textit{Japan}, \textit{Iran}, and \textit{Italy} rose and levelled off (Figure~\ref{fig:pandemic_korea}). For the Europe sub-topic (\textit{Italy-Spain-France-Germany}), the publication volumes rose and have kept rising. Lastly, the \textit{India} sub-topic shows publication volumes which are just beginning to rise. Using topic modelling on large corpora, combined with trend analysis, we can see how scientists have followed the progress of the pandemic through their publications. The interface, moreover, can help predict the trends in other countries and discover exciting movements in the relevance of other various research topics of the pandemic.

\section{Conclusion}

In this paper, we have presented one of the first works combining analysis and visualisation of large-volume literature datasets to highlight the impact of COVID-19 on many research communities. We show that it is possible to integrate advanced statistical topic modelling techniques into a visualisation pipeline which quickly: (a) abstracts thousands of publication entries into smaller themes; (b) extracts trend information; and (c) produces at-a-glance semantic visual overviews of rapidly changing corpora. This method, its techniques and interfaces, can help scientists browse, search, and access knowledge faster, and stay abreast of evolving themes.\newline

We have presented analysis, using topic and visual information, through different themes to summarise interesting aspects of the information inside the large volume of research literature. This analysis highlights: (a) the development of research regarding social distancing for the first time in 70 years; (b) insights into cross-domain initiatives to understand the consequences of this unprecedented situation; (c) the evolution in medical topics; and (d) the unfolding of the pandemic through publications. We hope the methods and findings may be useful as a reference guide for similar systems, to stimulate new ideas and directions of research, and to help in the fight against this pandemic.
\newline

\section*{Resources}

The datasets used in this paper can be found here:
\begin{itemize}
    \item \href{https://dimensions.figshare.com/articles/Dimensions\_COVID-19\_publications\_datasets\_and\_clinical\_trials/11961063}{https://dimensions.figshare.com/articles/Dimensions\_COVID-19\_publications\_datasets\_and\_clinical\_trials/11961063}
    \item \href{https://www.kaggle.com/allen-institute-for-ai/CORD-19-research-challenge}{https://www.kaggle.com/allen-institute-for-ai/CORD-19-research-challenge}
\end{itemize}

\noindent We have made available the visualisation interfaces at the following addresses:
\begin{itemize}
    \item \href{http://strategicfutures.org/TopicMaps/COVID-19/dimensions.html}{http://strategicfutures.org/TopicMaps/COVID-19/dimensions.html}
    \item \href{http://strategicfutures.org/TopicMaps/COVID-19/cord19.html}{http://strategicfutures.org/TopicMaps/COVID-19/cord19.html}
\end{itemize}

\section*{Acknowledgements}
This work was funded by the ORCA Hub (EPSRC grant: EP/R026173/1, website: \href{https://orcahub.org/}{orcahub.org}) and the \textit{Exploiting Impact Using a Modular Decision-Making Toolset} project (EPSRC Impact Acceleration Account: EP/R511535/1).

\newpage

\bibliographystyle{ACM-Reference-Format}
\bibliography{main}

\end{document}